\pdfoutput=1
\documentclass{sig-alternate-10pt}
\usepackage[utf8]{inputenc}
\usepackage{url}

\begin{document}

\title{A Smart Home is No Castle:\\Privacy Vulnerabilities of Encrypted IoT Traffic}
\numberofauthors{3}
\author{
	\alignauthor
	Noah Apthorpe\\
    	   \affaddr{Computer Science Dept.}\\
    	   \affaddr{Princeton University}\\
    	   \affaddr{apthorpe@cs.princeton.edu}
	\alignauthor 
	Dillon Reisman\\
    	   \affaddr{Computer Science Dept.}\\
    	   \affaddr{Princeton University}\\
    	   \affaddr{dreisman@princeton.edu}
	 \and
	\alignauthor
	 Nick Feamster\\
    	   \affaddr{Computer Science Dept.}\\
    	   \affaddr{Princeton University}\\
    	   \affaddr{feamster@cs.princeton.edu}  
}

\maketitle

\begin{abstract}
The increasing popularity of specialized Internet-connected devices and appliances, dubbed the Internet-of-Things (IoT), promises both new conveniences and new privacy concerns. 
Unlike traditional web browsers, many IoT devices have al\-ways-on sensors that constantly monitor fine-grained details of users' physical environments and influence the devices' network communications. 
Passive network observers, such as Internet service providers, could potentially analyze IoT network traffic to infer sensitive details about users. 
Here, we examine four IoT smart home devices (a Sense sleep monitor, a Nest Cam Indoor security camera, a WeMo switch, and an Amazon Echo) and find that their network traffic rates can reveal potentially sensitive user interactions even when the traffic is encrypted.
These results indicate that a technological solution is needed to protect IoT device owner privacy, and that IoT-specific concerns must be considered in the ongoing policy debate around ISP data collection and usage.
\end{abstract}

\section{Introduction}
\label{sec:intro}
The rapidly expanding availability and diversity of Internet of Things (IoT) devices for consumer smart homes promises to revolutionize how we interact with our living spaces.  
However, smart homes containing many Internet-connected devices raise substantial privacy concerns. 

The contents, patterns, and metadata of network traffic can all reveal sensitive information about a user's online activity.
In the past, online activity was primarily  limited to web browsing.
Unlike web browsers, smart home devices' always-on sensors transmit information about a user's offline activities on the Internet. 
This detailed data could be valuable in many contexts, including advertising and business intelligence.

We define a ``smart home device" as any single-purpose Internet-connected device intended for home use (e.g.\ a  thermostat, outlet, or blood-pressure monitor) \textit{or} a hub-like device that connects and controls multiple single-purpose devices (e.g.\ a Samsung SmartThings hub or Amazon Echo).
Interacting with a smart home device -- or even simply living within the range of a smart home device's environmental sensors -- changes how the device communicates with remote servers.
It would be a  privacy concern if a passive network observer, such as an Internet service provider (ISP), were able to infer user behavior from these changes in Internet traffic. 

Concerns over the abilities of network observers have led regulators to institute new rules on ISP data collection and usage  \cite{fcc-pr, fcc-nprm}.
Some opponents argue that stronger regulation is unnecessary, because the increasing pervasiveness of encryption prevents ISPs from observing sensitive data in traffic content \cite{swire}. 
Privacy advocates, however, argue that metadata and traffic patterns can reveal sensitive information even when traffic content is unavailable \cite{eff}. 
Further research detailing privacy vulnerabilities of encrypted traffic and metadata from IoT devices can help inform future regulation as IoT devices become more prevalent. 

In this paper, we develop a strategy that a passive network observer could use to infer consumer behavior from rates of IoT device traffic, even when the traffic is encrypted. 
This strategy relies on the limited-purpose nature of IoT devices to map traffic patterns to device states. 
We set up a smart home laboratory with a passive network tap to model how a real-life observer could collect traffic from an actual smart home.

We examine four commercially available smart home devices:  a Sense sleep monitor \cite{sense}, a Nest Cam Indoor security camera \cite{nestcam}, a Belkin WeMo switch \cite{wemo}, and an Amazon Echo \cite{echo}.  Using traffic from the Sense, a network observer could infer a user's sleeping patterns. Using traffic from the Nest Cam, an observer can infer when a user is actively monitoring the camera feed or when the camera detects motion in its field of vision.  Using traffic from the WeMo switch, an observer can detect when a physical appliance in a smart home is turned on or off. Using traffic from the Echo, an observer can detect when a user is interacting with an intelligent personal assistant.   

In light of these results, we are working to develop solutions that allow users to protect themselves from smart home privacy vulnerabilities. While additional research is necessary, the case studies we have performed motivate components of a general solution.  Traffic shaping will be required to mask the true rate of devices' network traffic.  VPN tunneling or another method of obfuscating packet headers would make it more difficult to identify individual devices. The challenge will be in combining multiple defensive strategies into a user-friendly and easily-deployable solution for smart homes.

\section{Threat Model}
\label{sec:threat-model}
Our privacy analysis assumes a passive network threat model with capabilities similar to an ISP.  
Specifically, an adversary in this model can observe and record all wide-area network traffic, including traffic to and from home gateway routers. 
The adversary cannot view local-area network traffic between devices behind a gateway router.
The adversary also cannot manipulate network traffic.
We assume that ISPs are typically uninterested in performing targeted active attacks on individual users.

In this paper, we do not use packet contents. 
In fact, we note that all 4 tested IoT devices use TLS/SSL when communicating with 1st and 3rd party cloud servers.
Our analysis relies entirely on metadata. 
IP packet headers, TCP packet headers, and send/receive rates are all available to the adversary.
This sort of metadata is regularly collected by major ISPs for traffic analysis.

Finally, we assume that the adversary can obtain and analyze IoT devices. For instance, using their own smart home laboratory,  an adversary can collect traffic data to help identify devices in live consumer traffic.

While we are only focused on the passive network threat model, other threat models (e.g.\ compromised home devices or Wi-fi eavesdroppers) could provide interesting opportunities for future study.

\section{Laboratory Smart Home}
\label{sec:setup}
We have set up a laboratory smart home environment to examine the network behavior of several on-market IoT devices. 

We included the following popular IoT devices in our laboratory smart home, covering a range of device types, manufacturers, and privacy concerns: 
1) Sense sleep monitor 
2) Nest Cam Indoor security camera
3) WeMo switch (smart power outlet)
4) Amazon Echo

We configured a Raspberry Pi 3 Model B  as a 802.11n wireless access point for use as the laboratory smart home's gateway router.
The Raspberry Pi 3 has a built-in WiFi antenna, so no additional hardware was needed. 
The Raspberry Pi ran the Raspbian Jessie OS, a version of Debian Linux optimized for the Raspberry Pi platform.  
Hostapd (host access point daemon) enabled mac80211-compliant access point and authentication server services.  
Dnsmasq  enabled DNS and DHCP services.  
An iptables NAT connected the Raspberry Pi's wireless interface to the wired interface connected by Ethernet to the WAN.  

This setup allowed us to record all packets to and from IoT devices connected to the Raspberry Pi.  
We recorded traffic from all devices operating concurrently and performed several controlled experiments with individual devices.
The resulting packet capture files were the raw data used for analysis in Sections~\ref{sec:strategy}~\&~\ref{sec:device-cs}.

\section{I\MakeLowercase{o}T Traffic Analysis Strategy}
\label{sec:strategy}
In this section we present a three step strategy that a passive network observer could use to identify devices in a smart home and infer user behavior. 

\subsection{Separate traffic into packet streams}
An adversary must first divide recorded network traffic into meaningful streams that can be used for further analysis.
In most standard consumer use cases, the home gateway router acts as a network address translator (NAT), rewriting local IP addresses of individual devices to a single public IP address given to the router by the ISP. 
This prevents an adversary from using IP addresses to divide traffic into per-device packet sets. 

Identifying and counting distinct clients behind a NAT is a known problem \cite{bellovin, kohno}.
However, it is always possible to separate network traffic into streams by the external IP address of the server communicating with the devices (``service IP") and, in cases where multiple devices use the same service IP, the TCP port rewritten by the NAT. 
While the devices we studied often communicate with multiple service IPs, we discovered that the adversary typically only needs to identify a single stream that encodes the device state.

\subsection{Label streams by type of device}

Once individual streams have been separated, the adversary next identifies what IoT device most likely is responsible for each stream.  Knowing what devices a consumer owns can be a serious privacy violation by itself. For example, a consumer might not want an ISP knowing they own an IoT blood sugar monitor or pacemaker.

In our case studies, the DNS queries associated with each stream could be mapped to a particular device (Figure \ref{fig:dns-queries}).  
For example, the Nest Cam queried domains from dropcam.com (the predecessor to the Nest Cam), while the Sense sleep monitor queried domains from hello.is (the company that makes the Sense).
An adversary could use a laboratory setup like our own to learn these mappings or perform reverse DNS lookups to pair service IPs with device-identifying domain names.

However, multiple devices from the same manufacturer might communicate with the same service IPs, making device identification using DNS more difficult. 
For example, the Belkin WeMo switch queried domains that could have been used by any type of Belkin device. 
Measuring the extent of this problem and finding solutions will be the subject of future study.

\subsection{Examine traffic rates}

Once an adversary identifies packet streams for a particular device, one or more of the streams are likely to encode device state.
Simply plotting send/receive rates of the streams (bytes per second) revealed potentially private user interactions for each device we tested.

An adversary with a laboratory smart home of their own can easily correlate variations in traffic rates with known user interactions. They can then map similar variations from live traffic to user behavior.
 
Even without a laboratory smart home, an adversary can still infer user interactions from traffic variations if they have identified the device and know its limited purpose.
For example, the Sense sleep monitor was both easily identified from DNS queries, and has a limited purpose.  A traffic spike from the monitor in the late evening, for instance, likely corresponds to when the user went to sleep.

\begin{figure}[t]
\begin{center}
\small
\begin{tabular}{ll}
\hline
\textbf{Device} & \textbf{DNS Queries} \\
\hline
Sense Sleep Monitor & \url{hello-audio.s3.amazonaws.com} \\
                                  & \url{hello-firmware.s3.amazonaws.com} \\
                                  & \url{messeji.hello.is} \\
                                  & \url{ntp.hello.is} \\
                                  & \url{sense-in.hello.is} \\
                                  & \url{time.hello.is} \\
\hline
Nest Security Camera & \url{nexus.dropcam.com} \\
                                    & \url{oculus519-vir.dropcam.com} \\
                                    & \url{pool.ntp.org} \\
\hline
WeMo Switch & \url{prod1-fs-xbcs-net-1101221371.}\\
                       &\quad \url{us-east-1.elb.amazonaws.com} \\
                       & \url{prod1-api-xbcs-net-889336557.} \\
                       &\quad \url{us-east-1.elb.amazonaws.com} \\
\hline
Amazon Echo & \url{ash2-accesspoint-a92.ap.spotify.com} \\
                       & \url{audio-ec.spotify.com} \\
                       & \url{device-metrics-us.amazon.com} \\
                       & \url{ntp.amazon.com} \\
                       & \url{pindorama.amazon.com} \\
                       & \url{softwareupdates.amazon.com} \\
\hline
\end{tabular}
\caption{DNS queries made by tested IoT devices during a representative packet capture.  Many queries can be easily mapped to a specific device or manufacturer.}
\label{fig:dns-queries}
\end{center}
\end{figure}

\section{Device Case Studies}
\label{sec:device-cs}
For all tested IoT devices, send/receive rates were sufficient for identifying user behaviors and interactions. Though all of the devices encrypted their traffic, encryption alone did not prevent privacy vulnerabilities.

\subsection{Sense sleep monitor}
\label{sec:sense}
Traffic to and from the Sense sleep monitor was easy to identify from DNS queries because all of the domains contain ``sense" or ``hello."  
Figure~\ref{fig:cs}A shows send/receive rates from the Sense over an approximately 12 hour period from 10:40pm to 10:40am.  
Notably, the send/receive rate peaked at times corresponding with user activity.  
The user shut off the light in the laboratory smart home and went to bed at 12:30am, temporarily got out of bed at 6:30am, and got out of bed in the morning at 9:15am. 
The traffic peaks correlated with these activities were not coincidental to this recording.  
Additional overnight traffic recordings also contained easily noticeable peaks when the user got into and out of bed.  

We believe that the ability of an network observer to tell when a user is sleeping, or at least in bed, from network send/receive rates constitutes a significant privacy vulnerability.  
ISPs can already guess when users are sleeping when network traffic from smartphones or PC web browsers decreases at night; 
however, this relies on many assumptions, e.g.\ that users only stop using their other devices immediately prior to sleeping, 
that everyone in the home sleeps at the same time and does not share other devices, 
and that users do not leave their other devices running to perform network-intensive tasks or updates while they sleep. 
The single-purpose IoT nature of the Sense sleep monitor makes  none of these assumptions necessary to infer users' sleeping patterns from Sense traffic.  

\subsection{Nest Cam Indoor security camera}
\label{sec:nestcam}
Our observations of the Nest camera indicate that it has at least two primary modes of operation: a live streaming mode and a motion detection mode.   
In the live streaming mode, the camera's video feed is either being actively viewed by the user through the Nest web/mobile application or the feed is being uploaded in real time to be stored on the cloud (for users with paid accounts).  
In the motion detection mode, the video stream is not being uploaded, but the camera is monitoring the stream locally for movement. 
If movement is observed, the camera records a snapshot of the video and alerts the user.  

Nearly all TCP traffic to and from the Nest camera is with dropcam.com domains, making it easy to identify from DNS queries.  
Figure~\ref{fig:cs}B shows send/receive rates from the Nest camera alternating between live streaming and motion detection mode every 2 minutes. 
The traffic rate is orders of magnitude higher in live streaming mode (and a short time afterward until the camera is notified that the user has stopped viewing the stream), allowing an adversary to easily determine whether or not the camera's live feed is being actively viewed or recorded. 

Figure~\ref{fig:cs}C shows that an adversary could also easily determine when a Nest camera detects movement when it is in motion detection mode. 
The camera was pointed at a white screen with a black square that changed location every two minutes.  These simulated motion events triggered clearly observable spikes in network traffic.  
This predictable variability in network send/receive rates would allow an adversary to observe the presence and frequency of motion inside a smart home.  

These issues are significant privacy vulnerabilities and physical security risks even though the content of the video stream remains protected by encryption.
It should not be possible for a third party to be able to determine when a security camera detects movement or is being actively monitored. 

\subsection{WeMo switch}
\label{sec:wemo}
The WeMo switch is an smart outlet controlled by a physical button on the device or through the WeMo smartphone app.  
WeMo switch traffic was more difficult to distinguish using DNS queries because all resolved addresses were from  ---xbcs---.amazonaws.com domains generic to Belkin. 
Nevertheless, the Belkin switch traffic was unique amongst the devices we tested for its regularity.  
The switch only has two states, on and off, and the network send/receive rates reflect this binarity.  
Figure~\ref{fig:cs}D shows WeMo network behavior when the switch is turned alternatively on and off every 2 minutes using the WeMo smartphone app.  
The spike in traffic every time the switch changes state clearly reveals user interactions with the device to an network observer.  

Additional recordings performed while the switch was turned on and off with the physical button on the device were effectively equivalent to Figure~\ref{fig:cs}D.  
This was initially surprising because there is no need for the device to contact the cloud in order to turn on or off in response to a physical button press. 
However, we realized that the state change is still communicated so the smartphone app can display the correct ``on" or ``off" icon for the device. 

While the WeMo switch send receive rates reveal user interactions with the device, they do not by themselves indicate whether the switch is on or off.  
The seriousness of this privacy vulnerability is debatable; however, it is a case where a network observer can learn that a human has turned power on or off to physical appliance. 
If combined with techniques of learning what type of appliance is plugged into the switch, this could have serious privacy implications.

\subsection{Amazon Echo}
\label{sec:echo}
\begin{figure}
\includegraphics[width=0.49\textwidth]{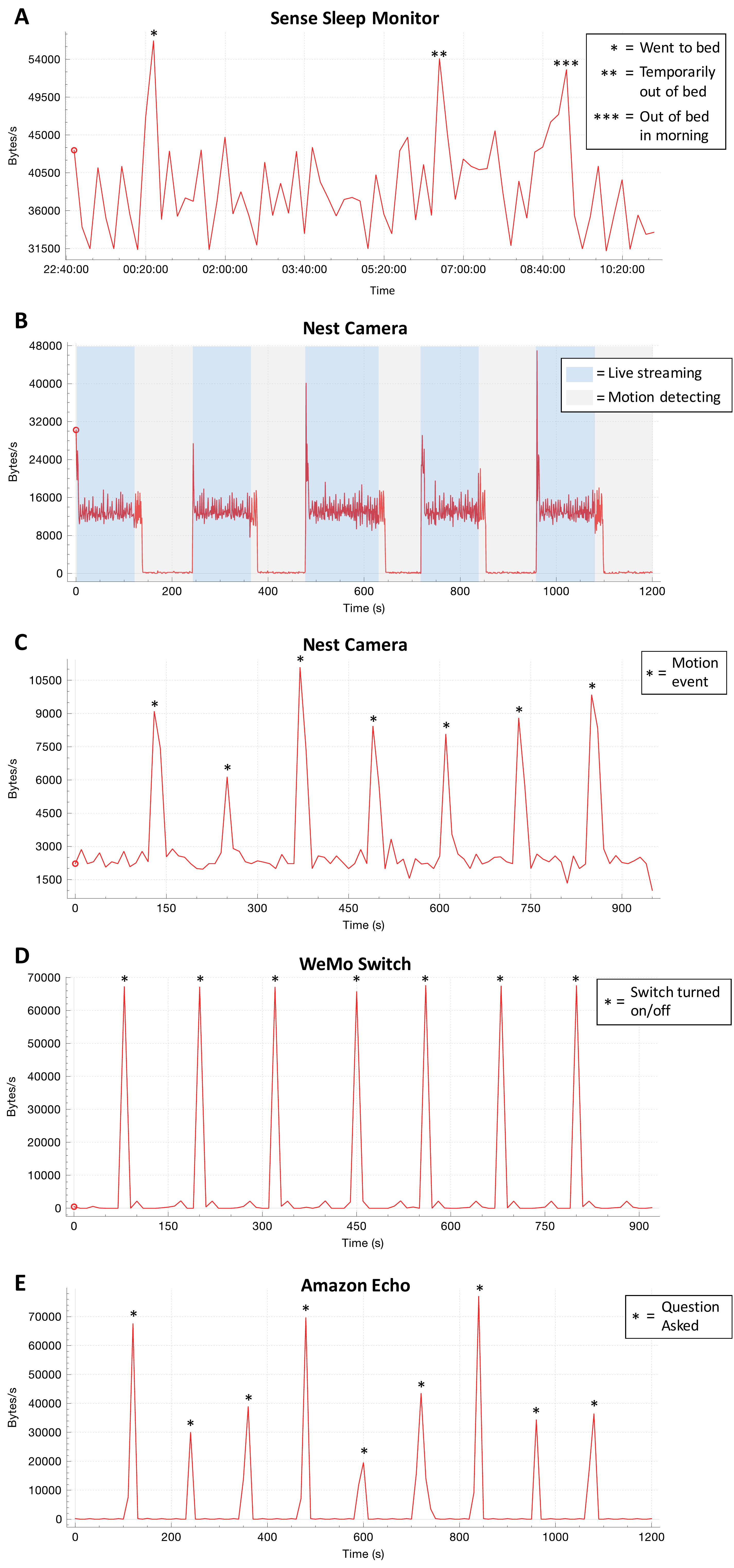}\\
\caption{Network traffic send/receive rates of selected IP streams from 4 commercially-available IoT devices during controlled experiments.  Clearly visible increases in send/receive rates directly correspond with user interactions.}
\label{fig:cs}
\end{figure}

The Amazon Echo is the most feature-diverse of the IoT devices we tested, and it had the correspondingly most complex network traffic profile.  
DNS queries from the Echo requested domains from amazon.com and spotify.com (for third-party service integration).  
This makes it difficult to assign all IP streams from the Echo to the correct device, but this turns out to be unnecessary for observing user interactions with the device.  
We tested the Echo by asking a a series of 3 questions (''what is the weather?" ``what time is it?" and ``what is the distance to Paris?") repeated 3 times, one question every 2 minutes. 
Figure~\ref{fig:cs}E shows the send/receive rates of SSL traffic between the Echo and a single amazon.com IP address during the experiment.  
Although the Echo sent and received other TCP traffic to different domains during this time, only the shown SSL traffic was noticeably correlated with the user interactions.  
As long as a network-level attacker can identify that particular IP stream as originating from an Echo, the SSL traffic spikes clearly indicate when user interactions occurred. 

To some, this may not seem to be a privacy vulnerability because the contents of the questions are encrypted.  However, simply learning the times of day when customers interact with a particular device could have unwanted advertising implications. 

\section{Discussion}
\label{sec:discussion}
We were surprised by how easy it would be for a passive network observer to infer user behavior from encrypted smart home traffic.
ISPs already collect enough traffic rate information to perform the analysis we described.
Because IoT devices encode the physical world in network traffic, this presents a novel privacy threat to consumers.
Regulatory agencies should keep this new context in mind when making rules governing ISP data collection and usage.

Future research should make the distinction between IoT devices that sense and encode the physical world, versus devices that are essentially wrappers around traditional web browsers.
For instance, when a user asks questions to the Amazon Echo, the Echo is just acting as an audio interface to a web search engine.
While inferring this use of the Echo from traffic is interesting in its own right, more privacy sensitive behaviors could be learned from more specialized classes of devices.
This is especially true if those devices are related to healthcare or physical security.

The user interactions we analyzed in our case studies are directly related to the limited purpose of the device. 
For example, we concluded that traffic from a sleep monitor correlates to when a user sleeps.
Further research could allow an adversary to infer higher order behaviors, such as whether the user has a sleeping disorder.
This would require larger curated datasets with controlled experiments representing a wider range of user behaviors.
Machine learning and advanced statistical techniques could also play an important part in inferring higher order behaviors.
More complex behaviors could also be inferred through the combination of traffic from multiple devices.

We would like to reiterate that all of the analyses we performed required only send/receive rates of encrypted traffic to successfully identify user behavior.
No deep packet inspection is necessary.
A systematic solution for preserving consumer privacy would therefore require obfuscating or shaping all smart home traffic to mask variations that encode real world behavior.
Incorporating VPN tunneling or another method of masking packet headers would also make device identification more difficult.
Ideally a solution would not negatively impact IoT device performance, should respect data limits, and would not require modification of proprietary device software.
Designing and implementing such a solution is a primary goal of our ongoing research.

\section{Conclusion}
\label{sec:conclusion}
IoT devices for smart homes are becoming increasingly pervasive; however, the privacy concerns of owning many Internet connected devices with always-on environmental sensors remain insufficiently addressed. 

We analyzed four commercially-available smart home devices and found that network traffic rates of all devices revealed user activities, making it apparent that encryption alone does not provide adequate privacy protection for smart homes.  Given the generality of our traffic analysis strategy and the limited-purpose nature of most IoT devices, we would not be surprised if  many other currently available smart home devices  suffer similar privacy vulnerabilities.

We hope that consumers will become better aware of these privacy vulnerabilities and that tools will be developed to protect smart homes from passive network observers.  We are working toward a user-friendly solution for smart home owners that will prevent traffic rates and other metadata from revealing offline user activities.  While we hope that a technological solution will be practical and sufficient, improved regulation of ISPs and other passive network observers may also be necessary to offset the unique privacy challenges posed by IoT devices. 

\section{Acknowledgments}
Thanks to Arvind Narayanan, Nina Taft, and Caio Burgardt for their thoughts and for contributing to the discussions that focused this work.  This work was partially supported by a Google Faculty Research Award and NSF Awards CNS-1526353 and CNS-1539902.

\bibliographystyle{abbrv}
\bibliography{SmartHomeNoCastle}

\end{document}